\documentclass[aps,prl,showpacs]{revtex4}

\usepackage{amsmath}
\usepackage{amssymb}
\usepackage{graphicx}

\usepackage{epsfig}

\begin{document}

\title{General Theory of the Zitterbewegung }

\author{Gyula D\'avid
}
\affiliation{Department of Atomic Physics,
E{\"o}tv{\"o}s University
\\ H-1117 Budapest, P\'azm\'any P{\'e}ter s{\'e}t\'any 1/A, Hungary}

\author{ J\'ozsef Cserti
}
\affiliation{Department of Physics of Complex Systems,
E{\"o}tv{\"o}s University
\\ H-1117 Budapest, P\'azm\'any P{\'e}ter s{\'e}t\'any 1/A, Hungary}

\begin{abstract}

We derive a general and simple expression for the time-dependence of the position operator of a  multi-band Hamiltonian with arbitrary matrix elements depending only on the momentum of the quasi-particle.
Our result shows that in such systems the Zitterbewegung like term related to a trembling motion of the quasi-particle,
always appears in the position operator.
Moreover, the Zitterbewegung is, in general, a \emph{multi-frequency} oscillatory motion of the quasi-particle.
We derive a few different expressions for the amplitude of the oscillatory motion including that related to the Berry connection matrix.
We present several examples to demonstrate how general and versatile our result is.

\end{abstract}

\pacs{03.65.-w,71.70.Ej,73.22.Pr,03.65.Vf}

\maketitle

Schr\"odinger in his original paper has predicted a `trembling' or in other words a rapid oscillatory motion of
the center of the free wave packet for relativistic electron~\cite{Schrodinger:cikk}.
However, the Zitterbewegung is not strictly a relativistic effect~\cite{Feshbach_RevModPhys.30.24,Barut_PhysRevD_23_2454,Barut_PhysRevD.31.1386,Cannata_PhysRevB.44.8599,
PhysRevLett.93.043004}
but can be observed in spintronic systems as well~\cite{Schliemann_Loss_PhysRevLett.94.206801}.
This work has initiated many other works~\cite{Shen_PhysRevLett.95.187203,Bruder_zitter_PhysRevB.72.045353,
Nikolic_PhysRevB.72.075335,Zawadzki_PhysRevB.72.085217,Schliemann:085323,
brusheim:205307,Katsnelson_Zitter_min-cond:cikk,
Zawadzki-3:cikk,Schliemann_side-jump_ISI:000243895600066,PhysRevLett.98.253005,PhysRevA.76.041801,
Winkler_Ulrich_ISI:000246890900071,
Zawadzky_graphene_ISI:000251326800144,Schliemann_Loss_ISI:000248866900038,
Zhang_ISI:000254292300028,Vaishnav_ISI:000255117800018,Wang_ISI:000255457100201,Schliemann_ISI:000254543000075,
Maksimova_ISI:000259690800083,Loss_SO_ISI:000260574400091,
Vertesi_ISI:000261215400050,Maksimova_ISI:000262245400084,Tan_ISI:000262246400065,atomic_Zitter_EPL:cikk,rusin:045416} with an aim to demonstrate the appearance of the Zitterbewegung not only for relativistic Dirac electrons.
In these works a common feature is that the oscillatory motion of the free particles can be described only by one frequency.

In our previous work we showed that for a wide class of Hamiltonians related to for example spintronic systems
and graphene, the Zitterbewegung can be treated in a unified way~\cite{Cserti_Zitter-ISI:000242409000014}.
Here the basic idea was that the Hamiltonian of several systems can be mapped to that modelling the precession of a virtual spin in an effective magnetic field.
The coupled equations for this virtual spin precession and the orbital motion of the quasi-particle can easily be solved.
Thus, the Zitterbewegung is arising because the virtual spin and the orbital motion for the quasi-particle are coupled.
Our work suggests as natural question whether the phenomenon of the Zitterbewegung also arises
for an even more general Hamiltonian.

In the present work we extend the Zitterbewegung phenomena to a broader class of quantum Hamiltonians for free (quasi-) particles.
In particularly, we derive a general and simple expression for the time-dependence of the position operator $\textbf{x}(t)$ for a  multi-band Hamiltonian given by \begin{equation}\label{H-matrix:eq}
H =
\begin{pmatrix}
H_{11}(\textbf{p}) & H_{12}(\textbf{p}) & \ldots & H_{1n}(\textbf{p}) \\
H_{21}(\textbf{p}) & H_{22}(\textbf{p}) & \ldots & H_{2n}(\textbf{p}) \\
\vdots & \vdots & \ddots & \vdots \\
H_{n1}(\textbf{p}) & H_{n2}(\textbf{p}) & \ldots & H_{nn}(\textbf{p}) \\
\end{pmatrix},
\end{equation}
where each matrix element is a differentiable function of the momentum $\textbf{p}$ of the particle itself and
$n\ge 2$ is the number of degrees of freedom of the system.
From our general expression for the position operator $\textbf{x}(t)$ we shall show that
i) the Zitterbewegung always appears for systems given by the Hamiltonian (\ref{H-matrix:eq}),
i) for $n>2$ the Zitterbewegung is in fact a \emph{multi-component} oscillatory motion of the free quasi-particle, ii) for $n=2$ we recover the results obtained earlier in the above mentioned references.

To find the time dependence of the position operator $\textbf{x}(t)$ of the quasi-particle
in Heisenberg picture one needs to calculate
\begin{equation}\label{x-t_op:eq}
    \textbf{x}(t) = e^{\frac{i}{\hbar}\, H t} \, \textbf{x}(0)\, e^{-\frac{i}{\hbar}\,  H t},
\end{equation}
where $\textbf{x}(0)$ is the position operator at $t=0$, ie, it equals to the position operator
in Schr\"odinger picture.
Because the momentum operator $\textbf{p}$ is a constant of motion we now work in the subspace of the Hilbert space for which the momentum $\textbf{p}$ is fixed.
Calculating the right hand side of equation (\ref{x-t_op:eq}) the crucial step is to decompose the Hamiltonian (\ref{H-matrix:eq}) into a sum of projection operators:
$H  = \sum_a E_a Q_a $, where $E_a$ is the $a$th eigenvalue of the Hamilton operator at a given momentum $\textbf{p}$, and $Q_a$ are projection operators satisfying the following relations:
$Q_a Q_b = \delta_{a b} \, Q_a $ and $\sum_a Q_a  = I_n$, where $I_n$ is the $n \times n$ unit matrix.
The position operator at time $t=0$ in Schr\"odinger picture and in momentum representation is
$\textbf{x}(0) = i \hbar \frac{\partial}{\partial \textbf{p}}$.
Consider the operator $U=e^{-\frac{i}{\hbar}\, H t}$ which is only a function of
the momentum operator $\textbf{p}$.
Then equation (\ref{x-t_op:eq}) can be rewritten as
\begin{eqnarray}\label{xU:eq}
   \textbf{x}(t) &=& U^{-1} \, \textbf{x}(0)\,  U
   = U^{-1}\, \left[\, \textbf{x}(0), U \, \right]
    + U^{-1} U \, \textbf{x}(0) \nonumber \\
    &=&
 \textbf{x}(0)+ i\, \hbar \,  U^{-1}\, \frac{\partial U}{\partial \textbf{p}},
\end{eqnarray}
where we have made use the relation
$\left[\textbf{x}(0), F(\textbf{p}) \right] =
i \hbar \, \frac{\partial F(\textbf{p})}{\partial \textbf{p}}$.
Decomposition of the Hamiltonian (\ref{H-matrix:eq}) into a sum of projection operators makes possible to write that
 $ e^{\pm \frac{ i}{\hbar}\, H t} =  \sum_a e^{\pm\frac{ i}{\hbar}\, E_a t} Q_a$.
Now, substituting these operators into equation (\ref{xU:eq}) and using the orthogonality
relations $Q_a Q_b = \delta_{a b} \, Q_a $ it yields
the time dependence of  the position operator:
\begin{subequations}
\label{x_op-Q-main-shift:eq}
\begin{eqnarray}
 \textbf{x}(t)  &=&  \textbf{x}(0)   + \sum_{a}\, \textbf{Z}_{aa} +
 t \sum_a \textbf{V}_a \, Q_a +
 \nonumber \\
 && +
 \sum_{a}\sum_{b\neq a}\, e^{i\, \omega_{ab}\, t} \, \textbf{Z}_{ab},
 \quad \text{where} \label{VZ_def-Q-shift:eq}  \\
 \textbf{V}_a(\textbf{p})&=&  \frac{\partial E_a (\textbf{p})}{\partial \textbf{p}}, \quad
 \textbf{Z}_{ab}(\textbf{p})=  i \hbar \, Q_a \frac{\partial Q_b}{\partial \textbf{p}},
 \label{V-Z_op-Q-main-shift:eq}%
\end{eqnarray}%
\end{subequations}%
and $\omega_{ab}= \frac{E_a-E_b}{\hbar}$ are the so-called \emph{beating frequencies}.
Here we call $\textbf{V}_a$ as \emph{partial velocities} and
$\textbf{Z}_{ab}$ as \emph{Zitterbewegung amplitudes}.
This is our central result in this work.

The interpretation of the different terms in (\ref{x_op-Q-main-shift:eq}) is as
follows.
The first term is the initial position of the quasi-particle.
In contrast to the usual dynamics (for systems with one degree of freedom),
the second and the fourth term are entirely new.
The second term is a displacement of the position operator independent of time.
The third term describes the motion of the quasi-particle with constant velocity which is,
in general, not equal to any of the partial velocities $\textbf{V}_a$.
Finally, the Zitterbewegung stems from the last, oscillatory term which describes the oscillatory motion of the quasi-particle.
The phenomenon of Zitterbewegung is similar to the beating effect with different frequencies in the classical wave mechanics.
The Zitterbewegung is a direct consequence of the coupling of different energy eigenstates for systems with more than one degree of freedom.
These terms in $\textbf{x}(t)$ are inherent of the Zitterbewegung and are expressed via the projection operator $Q_a$ related to the given Hamiltonian.

Equation (\ref{x_op-Q-main-shift:eq}) is the most general form for describing the phenomenon of the Zitterbewegung.
Our result shows explicitly that the oscillatory motion (the last term in equation (\ref{VZ_def-Q-shift:eq})) is a superposition of individual oscillatory motions with frequencies corresponding to all possible differences of the energy eigenvalues of the Hamiltonian (\ref{H-matrix:eq}).
Thus in the most general case the Zitterbewegung describes a \emph{multi-frequency} oscillatory motion of the quasi-particle.
This multi-frequency behavior of the Zitterbewegung has first been shown by Winkler \emph{et al.} in Ref.~\onlinecite{Winkler_Ulrich_ISI:000246890900071} for two specific systems, namely for the Kane modell and Landau-Rashba Hamiltonian.
However, the most clear manifestation of this multi-frequency behavior of the Zitterbewegung for the general Hamiltonian~(\ref{H-matrix:eq}) can be seen only in our main result (\ref{x_op-Q-main-shift:eq}).
In general, the Zitterbewegung  cannot be described by only one frequency
(this is the case only for systems with two different eigenenergies) but all of the differences between the different energy eigenvalues corresponding to the beating frequencies appear in the time dependence of the position operator.

Sometimes in the explicit calculation of the position operator $\textbf{x}(t)$ it is more useful to use a different form for the Zitterbewegung amplitudes $\textbf{Z}_{ab}$ given by (\ref{x_op-Q-main-shift:eq}).
Taking the derivative of the Hamilton operator $H = \sum_c \, E_c Q_c$ and the orthogonality relation $Q_c Q_b = \delta_{cb}\,  Q_b $  with respect to the momentum $\textbf{p}$
one can easily show that (for details see Sec.~A in the Appendix)
\begin{equation}\label{Z_Qa-grad-Qb:eq}
   \textbf{Z}_{ab} = i \hbar \, \frac{Q_a \frac{\partial H}{\partial \textbf{p}}\, Q_b}{E_b-E_a},
\end{equation}
valid for $a \ne b$.
Thus in the calculation of the Zitterbewegung amplitudes instead of knowing the derivative of the projection operators with respect to the momentum one needs to take only the derivative of the Hamiltonian.
Another form of the position operator  $\textbf{x}(t)$ is given in the Appendix.

We now consider several examples to demonstrate how versatile our result is to study different systems known in the literature (for more details see the Appendix).
Regarding the Zitterbewegung most of the systems studied in the literature are described by a Hamiltonian
with only two different eigenvalues~\cite{Schrodinger:cikk,Feshbach_RevModPhys.30.24,Barut_PhysRevD_23_2454,
Barut_PhysRevD.31.1386,Cannata_PhysRevB.44.8599,PhysRevLett.93.043004,
Schliemann_Loss_PhysRevLett.94.206801,Shen_PhysRevLett.95.187203,Bruder_zitter_PhysRevB.72.045353,
Nikolic_PhysRevB.72.075335,
Zawadzki_PhysRevB.72.085217,Schliemann:085323,Cserti_Zitter-ISI:000242409000014,
brusheim:205307,Katsnelson_Zitter_min-cond:cikk,Zawadzki-3:cikk,Schliemann_side-jump_ISI:000243895600066,
PhysRevLett.98.253005,PhysRevA.76.041801,Winkler_Ulrich_ISI:000246890900071,
Zawadzky_graphene_ISI:000251326800144,Schliemann_Loss_ISI:000248866900038,
Zhang_ISI:000254292300028,Vaishnav_ISI:000255117800018,Wang_ISI:000255457100201,Schliemann_ISI:000254543000075,
Maksimova_ISI:000259690800083,Loss_SO_ISI:000260574400091,
Vertesi_ISI:000261215400050,Maksimova_ISI:000262245400084,Tan_ISI:000262246400065,atomic_Zitter_EPL:cikk,rusin:045416}.
In such systems either the Hamiltonian itself is a $2 \times 2$ matrix or the dimension of the Hilbert space is
more than $2$ but the eigenvalues are degenerate and the Hamiltonian has only two different eigenvalues.
Thus for such systems it is useful to derive an alternative form for the time dependence of the position operator given by Eq.~(\ref{x_op-Q-main-shift:eq}).
Now the Hamiltonian in terms of projectors reads $H  = E_+ Q_+ + E_- Q_-$,
where $Q_\pm $ are the projection operators satisfying the usual relations mentioned above
and $E_\pm $ are the two eigenvalues of $H$.
Introducing the operator  $T= Q_+ - Q_-$ 
it is obvious that $T^2 = I$.
Note that in the mathematical literature the operators satisfying this relation are called involutary operator related to the mirror image in geometry.
The Hamiltonian can be rewritten as $H = \varepsilon \, I + (\hbar \omega /2)\, T$,
where $\varepsilon=(E_+ + E_-)/2$ and $\omega = (E_+ - E_-)/\hbar$.
Moreover, it is clear that $Q_\pm = (I\pm T)/2$.
Then using Eq.~(\ref{x_op-Q-main-shift:eq}) one can easily show that
\begin{subequations}
\label{VZ-T-1freki:eq}
\begin{eqnarray}
  \textbf{x}(t) &=& \textbf{x}(0) +  \textbf{W}\, t  + \textbf{Z}(t),
  \quad \text{where}  \label{xt-T-1freki:eq}\\
  \textbf{W} &=&
  \frac{\partial \varepsilon}{\partial \textbf{p}}\, I
  + \frac{1}{2}\, \frac{\partial \hbar \omega}{\partial \textbf{p}}\, T , \\
  \textbf{Z}(t) &=& \frac{\hbar}{2}\, \sin \left(\omega t\right)  \, \frac{\partial T}{\partial \textbf{p}}
  + \frac{i \hbar}{2}\, (1-\cos (\omega t))\, T\frac{\partial T}{\partial \textbf{p}}.
\end{eqnarray}%
\end{subequations}
From this result it is clear that in the oscillatory part of $\textbf{x}(t)$ there is
only one frequency component.

Equation (\ref{VZ-T-1freki:eq}) can easily be applied to the original Schr\"odinger's Zitterbewegung
and we find the same results as that by Schr\"odinger (see Sec.~B of the Appendix).
Another example is the Luttinger Hamiltonian~\cite{Luttinger_PhysRev.102.1030,Murakami_PhysRevB.69.235206} given by
\begin{equation}\label{Luttinger_Ham:eq}
    H  = \frac{1}{2m}\, \left[ \left(\gamma_1 + \frac{5}{2}\, \gamma_2\right) \textbf{p}^2
    - 2 \gamma_2 {\left(\textbf{p} \textbf{S}\right)}^2\right],
\end{equation}
where $\textbf{p}= (p_x,p_y,p_z)$ is the vector of the momentum operators, $\textbf{S}= (S_x,S_y,S_z)$
represents the spin operator with spin $3/2$,  $m$ and $\gamma_{1,2}$ are parameters of the model.
Using Eq.~(\ref{VZ-T-1freki:eq}) the position operator for Luttinger Hamiltonian can easily be derived (for more details see Sec.~C of the Appendix):
\begin{widetext}
\begin{eqnarray}\label{x_op-Lutt-VEG:eq}
\textbf{x}(t) &=& \textbf{x}(0)
    + \left(\frac{\gamma_1 +\frac{5}{2}\, \gamma_2}{m}\, I_4
    - \frac{2\gamma_2}{m}\,\frac{{\left(\textbf{p}\textbf{S}\right)}^2}{\textbf{p}^2}\right)\,\textbf{p}\, t
    + \sin \left(\omega t\right)
    \left( \frac{\textbf{p}{\left(\textbf{p}\textbf{S}\right)}^2}{\textbf{p}^4}
    -\frac{\textbf{S}\left(\textbf{p}\textbf{S}\right)
    + \left(\textbf{p}\textbf{S}\right)\textbf{S}}{2\, \textbf{p}^2}\right)
    \nonumber \\
 && +  (1- \cos (\omega t))\,
    \frac{\left(\textbf{p} \times \textbf{S}\right)\, {\left(\textbf{p}\textbf{S}\right)}^2
    + 2 \left(\textbf{p}\textbf{S}\right) \left(\textbf{p} \times \textbf{S}\right)\,
    \left(\textbf{p}\textbf{S} \right)
    +{\left(\textbf{p}\textbf{S}\right)}^2 \, \left(\textbf{p} \times \textbf{S}\right)}{4\, \textbf{p}^4},
\end{eqnarray}%
\end{widetext}%
where
$\omega = E_+ - E_- = (2 \gamma_2/ m)\, \textbf{p}^2$.
Note that this result agrees with that obtained by Winkler \emph{et al.}
in Ref.~\onlinecite{Winkler_Ulrich_ISI:000246890900071}, and by J.~Schliemann in a private communication
using a direct calculation of the right hand side of equation (\ref{x-t_op:eq}).

We also consider a non-trivial example for the Zitterbewegung not known in the literature,
namely the Zitterbewegung for bilayer graphene.
The Hamiltonian for bilayer graphene in the four by four representation is given in Refs.~\cite{Novoselov_Hall:ref,mccann:086805}.
Including the trigonal warping~\cite{mccann:086805,koshino:245403,cserti:066802}
the position operator $\textbf{x}(t)$ is more cumbersome but its structure
and the steps of the derivation are similar to the case when the trigonal warping is omitted.
Therefore, we now neglect the trigonal warping.
The position operator $\textbf{x}(t)$ can be derived using Eq.~(\ref{x_op-Q-main-shift:eq})
but to obtain the Zitterbewegung amplitudes it is more effective to use Eq.~(\ref{Z_Qa-grad-Qb:eq}).
The results is quite lengthy thus here we only refer to Sec.~E of the Appendix.
This is a non-trivial example for the Zitterbewegung.
Since the Hamilton operator for bilayer graphene has four different eigenvalues, we have six values of the energy differences.
However out of these six values there are only four different ones.
Therefore, the number of beating frequencies is only four.
In this example it is clear that the oscillatory motion of the electron is a superposition of individual oscillatory motions with \emph{four} different frequencies.

Recently, for specific systems the connection between the Zitterbewegung and the Berry phase has been noticed and investigated by Vaishnav and Clark~\cite{Vaishnav_ISI:000255117800018},
and Englman and V\'ertesi~\cite{Vertesi_ISI:000261215400050}.
We now show that the oscillatory terms, ie, the Zitterbewegung amplitudes in the position operator have a
close relation to the \emph{Berry connection matrix} appearing in the expression of the well-known Berry phase~\cite{Berry_phase_120:cikk} even for a general Hamiltonian (\ref{H-matrix:eq}).
To this end we present another form for position operator in terms of the eigenvectors of the Hamiltonian.

The projection operator can be expressed via the eigenvectors $ | u_{a,s} (\textbf{p})\rangle$ of
the Hamiltonian operator:
$Q_a(\textbf{p})= \sum_s | u_{a,s} (\textbf{p})\rangle \langle u_{a,s} (\textbf{p})|$,
where $s$ denotes the different eigenvectors in a subspace with the same energy eigenvalue $E_a$.
Then equation (\ref{x_op-Q-main-shift:eq}) can be rewritten as
\begin{subequations}
 \label{x_A_Berry_con:eq}
\begin{eqnarray}\label{x-eigenvec:eq}
   \textbf{x}(t) &=& \textbf{x}(0)  +
   t \sum_k \, \textbf{V}_k \, | u_k (\textbf{p})\rangle \langle u_k (\textbf{p}) |
   \nonumber \\
   && \hspace{-8mm} + \sum_{k, l}\, \left(e^{i\, \omega_{lk}\, t}-1 \right) \,
   \textbf{A}_{kl}(\textbf{p}) \, | u_k (\textbf{p})\rangle \langle u_l (\textbf{p})|,
   \\
   \textbf{A}_{kl}(\textbf{p}) &=&
   i \hbar \, \langle u_k (\textbf{p})|\, \frac{\partial}{\partial \textbf{p}} \, | u_l (\textbf{p})\rangle
    \label{Berry_con_A:eq}.
\end{eqnarray}%
\end{subequations}%
Here $\textbf{A}_{kl}$ is the so-called \emph{Berry connection matrix}.
The index $k$ labels the eigenvectors of the Hamiltonian with taking into account their multiplicity.

For systems with precessing spin in an effective magnetic field it turns out that to study the Zitterbewegung
Eq.~(\ref{x_A_Berry_con:eq}) is more appropriate than Eq.~(\ref{x_op-Q-main-shift:eq}).
This is demonstrated in Sec.~D of the Appendix, where we find the same result
for the position operator as that we derived before using a different approach~\cite{Cserti_Zitter-ISI:000242409000014}.

So far we concentrate on the structure of the Zitterbewegung for general Hamiltonian~(\ref{H-matrix:eq}).
However, the observation of the Zitterbewegung experimentally is more difficult problem.
It is well-known that the spatial size of the trembling motion of the relativistic electron
predicted by Schr\"odinger is of the order of Compton wavelength, and  its frequency is far beyond the present experimental possibilities~\cite{Schrodinger:cikk}.
The experimental observation of the Zitterbewegung in the non-relativistic quantum regime such as
in semiconductors with spin-orbit couplings~\cite{Schliemann_Loss_PhysRevLett.94.206801} is much more promising.
For example,  Vaishnav and Clark~\cite{Vaishnav_ISI:000255117800018},
and Merkl \emph{et al.}~\cite{atomic_Zitter_EPL:cikk} have proposed an experiment for observing Zitterbewegung
using ultra cold atoms, while Rusin and Zawadzki have proposed an experiment for observing Zitterbewegung probed by femtosecond laser pulses in graphene~\cite{rusin:045416}.
Very recently, Gerritsma \emph{et al.} have performed a quantum simulation of the Dirac equation using a single trapped ion and observed the Zitterbewegung~\cite{Ross_Nature_463_68_2010.cikk}.

One of the difficulty of observing the Zitterbewegung is the lack of time resolved probes.
The initial state, in general, is a superposition of different momentum-eigenstates:
$| \,\Psi_0 \rangle  = \int d^3 \textbf{p}\, \sum_k c_k(\textbf{p})\,  |u_k(\textbf{p})\rangle $.
Therefore, the expectation value is
$\bar{\textbf{x}}(t) = \langle \Psi_0 \,|\, \textbf{x}(t)\, | \, \Psi_0 \rangle
= \int d^3 \textbf{p}\, \sum_{k,l} \, c_k^*(\textbf{p}) \, c_l(\textbf{p})\,
     \langle u_k(\textbf{p}) \,|\, \textbf{x}(t) \, |u_l(\textbf{p}) \rangle $ which involves the integration over
the momentum $\textbf{p}$.
Since in Eq.~(\ref{x_op-Q-main-shift:eq}) the beating frequencies $\omega_{kl}(\textbf{p})$
depend on the momentum $\textbf{p}$ the integration over the momentum $\textbf{p}$ in $\bar{\textbf{x}}(t)$
may result in a strong suppression of the Zitterbewegung in time.
This problem can be circumvented if at least one beating frequency is independent of the momentum $\textbf{p}$ of the quasi-particle.
This is the case, for example, for bilayer graphene as shown in Sec.~E of the Appendix.
One beating frequency is constant and in the detectable regime $\omega = \gamma_1/\hbar \sim 0.6 \, {\textrm{fs}}^{-1} $, where $\gamma_1 \sim 0.4$ eV is the strongest interlayer coupling between two carbon atoms that are on the top of each other~\cite{Novoselov_Hall:ref,mccann:086805,mccann:086805,koshino:245403,cserti:066802}.
The amplitudes of the trembling motion will be investigated in the near future.
Our main aim in this paper is to establish a general theory for the Zitterbewegung.
On the other hand our general theory can be a good starting
point to search for systems that are realistic for experimental observation of the Zitterbewegung.

\emph{Conclusions.}---We presented a general theory for Zitterbewegung and derived a general and simple expression for the position operator $\textbf{x}(t)$ in Heisenberg picture
and in momentum representation, and for a given system it can easily be calculated.
In contrast to systems studied in the literature~\cite{Schrodinger:cikk,Feshbach_RevModPhys.30.24,Barut_PhysRevD_23_2454,
Barut_PhysRevD.31.1386,Cannata_PhysRevB.44.8599,
Schliemann_Loss_PhysRevLett.94.206801,Shen_PhysRevLett.95.187203,Bruder_zitter_PhysRevB.72.045353,
Nikolic_PhysRevB.72.075335,
Zawadzki_PhysRevB.72.085217,Schliemann:085323,Cserti_Zitter-ISI:000242409000014,
brusheim:205307,Katsnelson_Zitter_min-cond:cikk,Zawadzki-3:cikk,Schliemann_side-jump_ISI:000243895600066,Winkler_Ulrich_ISI:000246890900071,
Zawadzky_graphene_ISI:000251326800144,Schliemann_Loss_ISI:000248866900038,
Zhang_ISI:000254292300028,Vaishnav_ISI:000255117800018,Wang_ISI:000255457100201,Schliemann_ISI:000254543000075,
Maksimova_ISI:000259690800083,Loss_SO_ISI:000260574400091,
Vertesi_ISI:000261215400050,Maksimova_ISI:000262245400084,Tan_ISI:000262246400065,atomic_Zitter_EPL:cikk,rusin:045416}
the Zitterbewegung is a universal phenomenon and it always appears in the quantum dynamics of a system of quasi-particle with more than one degree of freedom.
Our main result~(\ref{x_op-Q-main-shift:eq}) shows that the Zitterbewegung, in general, is a multi-frequency beating effect in Heisenberg picture and has a close relation to the Berry connection.
We believe that our work presented here provides a better understanding and experimental guide for the Zitterbewegung studied intensively in the literature.

\emph{Acknowledgements:}---We acknowledge fruitful discussions with J. Schliemann, U. Z\"ulicke, T. Geszti and G. Sz\'echenyi.
Supported by the Hungarian Science Foundation OTKA under the contracts No. T48782 and 75529.

\bibliographystyle{c:/cserti/programok/tex/prsty}




\section{Appendix}

We now apply our main result given in the main text for Zitterbewegung to several systems known in the literature and also calculate the operator $\textbf{x}(t)$ for non-trivial systems previously not studied in the context of Zitterbewegung.
Below we also present the details of the derivation of the results mentioned in the main text.
Our examples show how powerful our general expressions (\ref{x_op-Q-main-shift:eq}) or (\ref{x_A_Berry_con:eq})
are for studying the phenomenon of Zitterwbewegung.


\subsection{A. Other equivalent forms for the position operator  }
\label{x-op-other:sec}

In this section we present the derivation of the result given by Eq.~(\ref{Z_Qa-grad-Qb:eq})
for the position operator $\textbf{x}(t)$.
Using the relation $i \hbar \,\sum_a \sum_b Q_a \, \frac{\partial Q_b}{\partial \textbf{p}} = 0$
following from $\sum_b Q_b =I_n$, Eq.~(\ref{x_op-Q-main-shift:eq}) can easily be rewritten as
\begin{eqnarray}\label{x-sol-exp-H-1:eq}
    \textbf{x}(t) - \textbf{x}(0) &=&
    t\, \sum_a \frac{\partial E_a }{\partial \textbf{p}}\, Q_a +
    i \hbar \, \sum_{a, b} \left(e^{i\omega_{ab}\, t}-1\right)\,
    Q_a\, \frac{\partial Q_b}{\partial \textbf{p}}.
\end{eqnarray}

We now present another form for the Zitterbewegung amplitudes $\textbf{Z}_{ab}$
given by  Eq.~(\ref{x_op-Q-main-shift:eq}).
Taking the derivative of the Hamilton operator $H = \sum_c \, E_c Q_c$ with respect to the momentum $\textbf{p}$ we have
\begin{equation}\label{Hder_p:eq}
    \frac{\partial H }{\partial \textbf{p}} = \sum_c \left(\frac{\partial E_c}{\partial \textbf{p}}\, Q_c
    + E_c \frac{\partial Q_c}{\partial \textbf{p}}\right).
\end{equation}
Similarly, from the orthogonality relation $Q_c Q_b = \delta_{cb}\,  Q_b $ we find that
\begin{equation}\label{orth-der:eq}
    \frac{\partial Q_c}{\partial \textbf{p}}\,  Q_b + Q_c \frac{\partial Q_b}{\partial \textbf{p}}
    = \delta_{cb}\, \frac{\partial Q_b}{\partial \textbf{p}}.
\end{equation}
Thus using Eqs.~(\ref{Hder_p:eq}) and (\ref{orth-der:eq}) for a given $a$ and $b$ we may write:
\begin{eqnarray}\label{aHderQ:eq}
    Q_a \, \frac{\partial H }{\partial \textbf{p}}\, Q_b &=&
    \sum_c \left(\frac{\partial E_c}{\partial \textbf{p}}\, Q_a Q_c Q_b
    + E_c \, Q_a \frac{\partial Q_c}{\partial \textbf{p}}\, Q_b\, \right) =
    \frac{\partial E_b}{\partial \textbf{p}}\, Q_a Q_b +
    \sum_c E_c Q_a \left(\delta_{cb}\, \frac{\partial Q_b}{\partial \textbf{p}} - Q_c \frac{\partial Q_b}{\partial \textbf{p}} \right)  \nonumber \\
     &=& \delta_{ab}\, \frac{\partial E_a}{\partial \textbf{p}}\,  Q_a +
     \left(E_b - E_a \right) \, Q_a\, \frac{\partial Q_b}{\partial \textbf{p}},
\end{eqnarray}
where we have made used of the orthogonality relations $Q_a Q_b =\delta_{ab}\, Q_a $.
Hence we find that for $a \neq b$
\begin{equation}\label{aHderQ-1:eq}
    Q_a\, \frac{\partial Q_b}{\partial \textbf{p}} =
    \frac{Q_a \, \frac{\partial H }{\partial \textbf{p}}\, Q_b}{E_b-E_a}.
\end{equation}
Now from this relation it is simple to obtain the alternative expression for the Zitterbewegung amplitudes given
by Eq.~(\ref{Z_Qa-grad-Qb:eq}).

To get deeper insight into the phenomenon of the Zitterbewegung it is useful to present another form for the time dependent operator $\textbf{x}(t)$.
For the given index $a$ and $b$ in Eq.~(\ref{x-sol-exp-H-1:eq}) we now introduce
the operator $T_{ab} = Q_a - Q_b$. Then it follows that $T^2_{ab} = Q_a + Q_b$ and
$T^3_{ab} = T_{ab}$.
The operators satisfying the latter equation are called weak involutary operators.
It is obvious that an involutary operator is always a weak involutary operator.
Note also that $T_{ba}= - T_{ab}$.
From the definition of the operator $T_{ab}$ we find that $Q_a=\frac{1}{2} \left(T_{ab}^2 + T_{ab}\right)$
and $Q_b=\frac{1}{2} \left(T_{ab}^2 - T_{ab}\right)$.
We now express the second and third terms in Eq.~(\ref{x-sol-exp-H-1:eq}) in terms of the operator $T_{ab}$.
Then one can easily show that
\begin{eqnarray}\label{VTshow:eq}
    \sum_{a,b} \frac{\partial \hbar \omega_{ab}}{\partial \textbf{p}} \, T_{ab} &=&
    - 2\frac{\partial \sum_a E_{a}}{\partial \textbf{p}}\, I_n
    + 2 n \, \sum_a \frac{\partial E_{a}}{\partial \textbf{p}}\, Q_a , \\
    Q_a \, \frac{\partial Q_{b}}{\partial \textbf{p}} + Q_b\, \frac{\partial Q_{a}}{\partial \textbf{p}}
    &=& \frac{1}{2}\, T^2_{ab}\, \frac{\partial T_{ab}}{\partial \textbf{p}}\, T_{ab}, \\
    Q_a \, \frac{\partial Q_{b}}{\partial \textbf{p}} - Q_b\, \frac{\partial Q_{a}}{\partial \textbf{p}}
    &=& \frac{1}{2}\, T_{ab}\, \frac{\partial T_{ab}}{\partial \textbf{p}}\, T_{ab},
\end{eqnarray}
where $I_n$ is the $n \times n$ unit matrix.
Then it is easy to rewrite Eq.~(\ref{x-sol-exp-H-1:eq}) as
\begin{subequations}\label{x_op-Tab_main:eq}
\begin{eqnarray}
\label{x_op-Tab:eq}
 \textbf{x}(t) &=& \textbf{x}(0) + \textbf{W} \, t  +
 \frac{i \hbar}{4} \, \sum_a \sum_{b\neq a }\, \left[
 \, i \, \sin(\omega_{ab}\, t) + \, T_{ab}\, \cos(\omega_{ab}\, t)
 \right]\,T_{ab}\, \frac{\partial T_{ab}}{\partial \textbf{p}}\, T_{ab},  \\
 &=& \textbf{x}(0) + \textbf{W} \, t  +
 \frac{i\hbar}{4}\, \sum_{a,b} \left(e^{i\, \omega_{ab}\, T_{ab}\, t}-1\right)\,
 \frac{\partial T_{ab}}{\partial \textbf{p}}\, T_{ab},
 \quad \text{where} \label{VZ_def_T:eq}  \\
 \textbf{W} &=& \frac{1}{n}\, \frac{\partial \sum_a E_a}{\partial \textbf{p}}\, I_n +
 \frac{1}{2n}\, \sum_{a,b} \frac{\partial \hbar \omega_{ab}}{\partial \textbf{p}} \, T_{ab}.
\end{eqnarray}%
\end{subequations}%
In our general case, when there are more than one frequency component in the oscillatory term owing to
the fact that the system has more than two different energy levels, the time dependent
form of the oscillatory term has the same exponential form between the corresponding energy levels
as in Schr\"odinger's case (see the next section of the Appendix).
In summary, Eqs.~(\ref{x-sol-exp-H-1:eq}) and (\ref{x_op-Tab_main:eq}) are the two alternative forms of our main result
(\ref{x_op-Q-main-shift:eq}) for the position operator.

\subsection{B. Revisiting Schr\"odinger's original derivation}
\label{Sch_revisit:sec}

Our expression given by Eq.~(\ref{VZ-T-1freki:eq})
can directly be used to study the Zitterbewegung for free electron described by the Dirac equation.
This problem was  originally examined by Schr\"odinger~\cite{Schrodinger:cikk}.
The Dirac Hamiltonian reads
\begin{equation}\label{Dirac:eq}
    H = c \, \mbox{\boldmath $\alpha$} \textbf{p} + m c^2 \beta,
\end{equation}
where $\mbox{\boldmath $\alpha$}$ and $\beta$ are the usual $4\times 4$ matrices satisfying the relations
$\alpha_1^2 = \alpha_2^2 = \alpha_3^2 = \beta^2 = 1$, $\alpha_i \beta + \beta \alpha_i = 0$ and
$\alpha_i \alpha_k + \alpha_k \alpha_i = 2 \delta_{ik}$.
One can show that $H^2 = E^2$, where $E=\sqrt{c^2 \textbf{p}^2 +{\left(m c^2\right)}^2}$.
Introducing the operator $T=H /E$ it is clear that $T$ is an involutory operator since $T^2 = H^2/E^2 =I$, where 
$I$ is a four by four unit matrix.
Then the Hamiltonian can be written as $H = \frac{\hbar \omega}{2} \, T$.
Thus, from Eq.~(\ref{VZ-T-1freki:eq}) the time dependence of the position operator is
\begin{equation}\label{x-t_Sch-1:eq}
    \textbf{x}(t) = \textbf{x}(0) +
    \frac{\hbar}{2}\, \frac{\partial \omega}{\partial \textbf{p}} \, T \, t +
    \frac{\hbar}{2}\, \sin(\omega t)\, \frac{\partial T}{\partial \textbf{p}} +
     \frac{\hbar}{2i}\,\left(\cos (\omega t)-1\right)\, T \frac{\partial T}{\partial \textbf{p}}.
\end{equation}
We need to evaluate the gradient of $\omega$ and $T$ with respect to $\textbf{p}$:
\begin{eqnarray}\label{grad-omega-Sch:eq}
    \frac{\hbar}{2}\, \frac{\partial \omega}{\partial \textbf{p}} &=& \frac{\partial E}{\partial \textbf{p}}
    =\frac{c^2 \textbf{p}}{E} \equiv \textbf{V} T,
    \quad \text{where} \quad \textbf{V}= c^2 \textbf{p} {H }^{-1}
    ,  \\
    \frac{\partial T}{\partial \textbf{p}} &=& \frac{\partial \left(H /E\right)}{\partial \textbf{p}} =
    \frac{\textbf{v}-\textbf{V}}{E}, \quad \text{where} \quad \textbf{v}=  c\, \mbox{\boldmath $\alpha$}.
    \label{grad-T-Sch:eq}
\end{eqnarray}
Here $\textbf{V}$ is the classical relativistic velocity of the particle with momentum $\textbf{p}$ and energy $H$.
Then $\textbf{x}(t)$ becomes
\begin{eqnarray}\label{x-t_Sch-2:eq}
    \textbf{x}(t) &=& \textbf{x}(0) + \textbf{V}\, t +
    \frac{\hbar}{2}\, \sin(\omega t)\, \frac{\textbf{v}-\textbf{V}}{E} +
    \frac{\hbar}{2i }\, \left(\cos (\omega t)-1\right)\, T \, \frac{\textbf{v}-\textbf{V}}{E} \nonumber \\
    &=& \textbf{x}(0) + \textbf{V}\, t + i \hbar\,
    \left(\textbf{v} - \textbf{V} \right) \frac{e^{-\frac{2 i H }{\hbar}\, t}-I}{2 H }.
\end{eqnarray}
This results agrees with that given by Schr\"odinger in his original work~\cite{Schrodinger:cikk}
on the Zitterbewegung of the relativistic free electron.

\subsection{C. Luttinger Hamiltonian }
\label{Luttinger:sec}

We now present a non-trivial example for calculating the Zitterbewegung in case of Luttinger Hamiltonian given by
\begin{equation}\label{Luttinger_Ham:eq}
    H  = \frac{1}{2m}\, \left[ \left(\gamma_1 + \frac{5}{2}\, \gamma_2\right) \textbf{p}^2
    - 2 \gamma_2 {\left(\textbf{p} \textbf{S}\right)}^2\right],
\end{equation}
where $\textbf{p}= (p_x,p_y,p_z)$ is the vector of the momentum operators and $\textbf{S}= (S_x,S_y,S_z)$
represents the spin operator with spin $3/2$, while $m$ and $\gamma_{1,2}$ are parameters of the model~\cite{Luttinger_PhysRev.102.1030,Murakami_PhysRevB.69.235206} (in this section we take $\hbar =1 $).
The Hamiltonian can be expressed in terms of the projection operators
$Q_L$ and $Q_H$ as~\cite{Murakami_PhysRevB.69.235206}
\begin{eqnarray}
  H &=& E_H(\textbf{p}) Q_H(\textbf{p})+E_L(\textbf{p}) Q_L(\textbf{p}),\,\, \text{where} \\
  Q_L(\textbf{p}) &=& \frac{9}{8}\, I_4-\frac{1}{2\, \textbf{p}^2}\, {\left(\textbf{p} \textbf{S}\right)}^2,
  \label{QL-def:eq}\\
  Q_H(\textbf{p}) &=& I_4-Q_L(\textbf{p}),
  \label{QH-def:eq}
\end{eqnarray}
and $I_4$ is the $4 \times 4$ unit matrix,
and the double degenerate eigenvalues are $E_L(\textbf{p})= \frac{\gamma_1 + 2\gamma_2}{2m}\, \textbf{p}^2$
and $E_H(\textbf{p})= \frac{\gamma_1 - 2\gamma_2}{2m}\, \textbf{p}^2$
corresponding to the light-hole (L) and the heavy-hole (H) bands.
The projection operators $Q_L$ and $Q_H$ satisfy the usual orthogonality relations.

To find the time dependence of the position operator we use Eq.~(\ref{VZ-T-1freki:eq}) valid for Hamiltonian with two eigenvalues.
Take $Q_L = Q_+$ and $Q_H = Q_-$ then we find $T= Q_+ - Q_-
= \frac{5}{4}\, I_4 - \frac{{\left(\textbf{p} \textbf{S}\right)}^2}{\textbf{p}^2}$,
$\varepsilon = \frac{E_+ + E_-}{2}= \frac{\gamma_1}{2 m}\, \textbf{p}^2$
and $\omega = E_+ - E_- = \frac{2 \gamma_2}{m}\, \textbf{p}^2$.
Furthermore, we need the derivative of $T$ with respect to $\textbf{p}$:
\begin{equation}\label{T-der-Luttinger;eq}
    \frac{\partial T}{\partial \textbf{p}} = \frac{2\, \textbf{p}{\left(\textbf{p}\textbf{S}\right)}^2}{\textbf{p}^4}
    -\frac{\textbf{S}\left(\textbf{p}\textbf{S}\right)+ \left(\textbf{p}\textbf{S}\right)\textbf{S}}{\textbf{p}^2}.
\end{equation}
Now the time dependence of the position operator reads
\begin{equation}\label{x_op-Lutt-1:eq}
   \textbf{x}(t) = \textbf{x}(0) +  \textbf{W}\, t  + \textbf{Z}(t),
\end{equation}
where $\textbf{W}$ and $\textbf{Z}(t)$ can easily be obtained from Eq.~(\ref{VZ-T-1freki:eq}):
\begin{eqnarray}
  \label{V-Luttinger:eq}
    \textbf{W} &=& \frac{\gamma_1 +\frac{5}{2}\, \gamma_2}{m}\, \textbf{p}\, I_4
    - \frac{2\gamma_2}{m}\,\frac{\textbf{p}{\left(\textbf{p}\textbf{S}\right)}^2}{\textbf{p}^2}, \\
  \textbf{Z}(t) &=& \frac{1}{2}\, \sin \left(\omega t\right)  \, \frac{\partial T}{\partial \textbf{p}}
  -   \frac{i }{4 \, \textbf{p}^4}\, (\cos (\omega t) - 1)
  \left[
  {\left(\textbf{p}\textbf{S}\right)}^2 \textbf{S}\left(\textbf{p}\textbf{S}\right)
  - \left(\textbf{p}\textbf{S}\right)\textbf{S}  {\left(\textbf{p}\textbf{S}\right)}^2
  +{\left(\textbf{p}\textbf{S}\right)}^3 \textbf{S}
  -\textbf{S}  {\left(\textbf{p}\textbf{S}\right)}^3
  \right].  \label{Zt-1:eq}
\end{eqnarray}
The expression $\textbf{Z}(t)$ can be further simplified using the following identity for the spin operators
(can be proven using the commutation relations $\left[S_j,S_k \right]= i \, \varepsilon_{jkl}S_l $):
\begin{equation}\label{S_ident:eq}
    \textbf{S} \left(\textbf{p}\textbf{S}\right)- \left(\textbf{p}\textbf{S}\right) \textbf{S} =
    i\, \textbf{p} \times \textbf{S}.
\end{equation}
Finally, the position operator (\ref{x_op-Lutt-1:eq}) for Luttinger Hamiltonian can be written as
\begin{eqnarray}\label{x_op-Lutt-VEG-2:eq}
\textbf{x}(t) &=& \textbf{x}(0)
    + \left(\frac{\gamma_1 +\frac{5}{2}\, \gamma_2}{m}\, I_4
    - \frac{2\gamma_2}{m}\,\frac{{\left(\textbf{p}\textbf{S}\right)}^2}{\textbf{p}^2}\right)\,\textbf{p}\, t
    + \sin \left(\omega t\right)
    \left( \frac{\textbf{p}{\left(\textbf{p}\textbf{S}\right)}^2}{\textbf{p}^4}
    -\frac{\textbf{S}\left(\textbf{p}\textbf{S}\right)
    + \left(\textbf{p}\textbf{S}\right)\textbf{S}}{2\, \textbf{p}^2}\right)
    \nonumber \\
 && +  (1- \cos (\omega t))\,
    \frac{\left(\textbf{p} \times \textbf{S}\right)\, {\left(\textbf{p}\textbf{S}\right)}^2
    + 2 \left(\textbf{p}\textbf{S}\right) \left(\textbf{p} \times \textbf{S}\right)\,
    \left(\textbf{p}\textbf{S}\right)
    +{\left(\textbf{p}\textbf{S}\right)}^2 \, \left(\textbf{p} \times \textbf{S}\right)}{4\, \textbf{p}^4}.
\end{eqnarray}
Note that this result agrees with that obtained by Winkler \emph{et al.}
in Ref.~\cite{Winkler_Ulrich_ISI:000246890900071}, and by J.~Schliemann in a private communication
using a direct calculation of the right hand side of Eq.~(\ref{x-t_op:eq}).

\subsection{D. Spin in an effective magnetic field}
\label{spin_x:sec}

In this section we consider the Zitterbewegung for systems mapped to a system of  virtual spin in an effective magnetic field.
Such classes of Hamiltonian have been previously studied by the present authors
in Ref.~\cite{Cserti_Zitter-ISI:000242409000014} using a different approach.
For many systems~\cite{Cannata_PhysRevB.44.8599,
Schliemann_Loss_PhysRevLett.94.206801,Shen_PhysRevLett.95.187203,Bruder_zitter_PhysRevB.72.045353,Nikolic_PhysRevB.72.075335,
Zawadzki_PhysRevB.72.085217,Schliemann:085323,Cserti_Zitter-ISI:000242409000014,
brusheim:205307,Katsnelson_Zitter_min-cond:cikk,Zawadzki-3:cikk,Schliemann_side-jump_ISI:000243895600066,Winkler_Ulrich_ISI:000246890900071,
Zawadzky_graphene_ISI:000251326800144,Schliemann_Loss_ISI:000248866900038,
Zhang_ISI:000254292300028,Vaishnav_ISI:000255117800018,Wang_ISI:000255457100201,Schliemann_ISI:000254543000075,
Maksimova_ISI:000259690800083,Loss_SO_ISI:000260574400091,
Vertesi_ISI:000261215400050,Maksimova_ISI:000262245400084,Tan_ISI:000262246400065,atomic_Zitter_EPL:cikk,rusin:045416}
the Hamiltonian can be written in a quite general form:
\begin{equation}
H= \varepsilon({\bf p})\, I + \mbox{\boldmath $\Omega$({\bf p})}\, {\bf S},
\label{spin-H:eq}
\end{equation}
where $I $ is the unit matrix in spin space, and the system is characterized by the one-particle energy
dispersion $\varepsilon({\bf p})$ and the effective magnetic field
$\mbox{\boldmath $\Omega$}({\bf p})$ coupled to a virtual spin ${\bf S}$ with magnitude $S$.
Here we assume that $\varepsilon({\bf p})$ and $\mbox{\boldmath $\Omega$}({\bf p})$
are differentiable functions of the momentum ${\bf p}=(p_x,p_y,p_z)$.
In Ref.~\cite{Cserti_Zitter-ISI:000242409000014} we listed a few systems (together with the
effective magnetic field $\mbox{\boldmath $\Omega$}({\bf p})$) that are currently intensely studied
in spintronics, and in the research of graphene and superconductors.
To find operator $\textbf{x}(t)$ for $S = 1/2$ one can use our result given by Eq.~(\ref{VZ-T-1freki:eq}),
however for $S \ne 1/2$ the derivation is more subtle.
We now highlight the main steps in this calculation  for the case of $S \ne 1/2$.

Now it is more appropriate to start from the alternative form for $\textbf{x}(t)$
given in the main text by Eq.~(\ref{x_A_Berry_con:eq}) that involves the Berry connection matrix.
Solving the eigenvalue problem for Hamiltonian (\ref{spin-H:eq}) we find that the eigen-energies are
$E_m = \varepsilon({\bf p}) + \hbar \, \Omega \, m$, where $\Omega = \sqrt{\mbox{\boldmath $\Omega$}^2} $ and
$m=-S,-S+1, \dots, S-1,S$ labels the eigenvectors $| m \rangle$ of $H $
which are the same as the eigenvectors of operator $S_z$, the component of the spin operator
$\textbf{S}$ with spin quantization axis $z$ pointing along vector $\mbox{\boldmath $\Omega$}$.
In what follows, it is useful to introduce the unit vector $\textbf{n} = \mbox{\boldmath $\Omega$}/\Omega$.
Then $(\textbf{n} \textbf{S}_0) |m \rangle = m | m \rangle$, where ${\bf S}_0$ is the spin operator ${\bf S}$
in Schr\"odinger picture, ie, it is time independent.
The Hamilton operator becomes
$H = \sum_{m=-S}^{m=S}\, E_m\, | m  \rangle \langle m |$.

In the second term of Eq.~(\ref{x-eigenvec:eq}) the velocity operator
$\textbf{W}= \sum_k \textbf{V}_k \, | u_k \rangle \langle u_k | $ can be obtained as
\begin{eqnarray}\label{V-spin:eq}
    \textbf{W} &=& \sum_{m=-S}^{m=S}\, \frac{\partial E_m}{\partial \textbf{p}}\, | m \rangle \langle m | =
    \frac{\partial \varepsilon}{\partial \textbf{p}}\, \sum_{m=-S}^{m=S}\, | m \rangle \langle m | +
    \frac{\partial \hbar\, \Omega}{\partial \textbf{p}}\, \sum_{m=-S}^{m=S}\, m \, | m \rangle \langle m |
    = \frac{\partial \varepsilon}{\partial \textbf{p}}\, I +
    \frac{\partial \hbar\, \Omega}{\partial \textbf{p}}\,(\textbf{n} \textbf{S}_0 ) ,
\end{eqnarray}
where we have made use of the fact that the eigenvectors $| m \rangle$ form a complete set.

Similarly, we can calculate the Berry connection matrix $\textbf{A}_{mn}$ in Eq.~(\ref{Berry_con_A:eq}).
Using the general relation
\begin{equation}\label{Berry_con_der-Ham:eq}
    \langle m |\, \frac{\partial}{\partial \textbf{p}} \, | n \rangle =
    \frac{\langle m |\, \frac{\partial H }{\partial \textbf{p}} \, | n \rangle}{E_n-E_m},
    \quad m\ne m^\prime ,
\end{equation}
the $m n$ matrix element of the $j$th component of the vector operator $\textbf{A}$ can be expressed as
\begin{equation}\label{Berry_spin-1:eq}
    {\left(A_j\right)}_{m n} = \frac{i \hbar}{E_n - E_m}\, \sum_{l=1,2,3}\,K_{jl} \langle m |\,  S_l | n \rangle ,
\end{equation}
valid for $m \neq n$ and where the matrix $\textbf{K}$ is defined as $K_{jk}= \partial \Omega_k /\partial p_j $.
Then employing the well-known relations between the spin operators $S_x, S_y$ and  $S_z$, one can evaluate
the matrix elements of the spin operator $S_j$ in the above equation.
After some algebra the time dependence of the position operator becomes
\begin{eqnarray}
\textbf{x}(t) &=& \textbf{x}(0) +  \textbf{W}\, t  + \textbf{Z}(t),
\quad \text{where} \label{x_op-spin:eq} \\
\textbf{W} &=& \frac{\partial \varepsilon}{\partial \textbf{p}}\, I +
    \frac{\partial \hbar\, \Omega}{\partial \textbf{p}}\,(\textbf{n} \textbf{S}_0 ) , \\
\textbf{Z}(t)&=& \frac{\sin \Omega t }{\Omega} \,
{\bf K} \left(I-{\bf n}\circ{\bf n} \right)\, {\bf S}_0
+ \frac{1-\cos \Omega t }{\Omega}\,
{\bf K} \left({\bf n} \times {\bf S}_0 \right),
\label{rH_t:eq}
\end{eqnarray}%
and ${\bf n}\circ{\bf n}$ denotes the outer or direct product, ie,
${({\bf n}\circ{\bf n})}_{jk} = n_j n_k$.
Note that this result agrees exactly with that derived with a different method in our previous publication\cite{Cserti_Zitter-ISI:000242409000014}.

\subsection{E. Bilayer graphene without trigonal warping }
\label{bilayer:sec}

We now present a non-trivial example not known in the literature, namely the bilayer graphene system.
In this case we show that the oscillatory motion of the electron is a superposition of individual oscillatory motions with four different frequencies.

The Hamiltonian for bilayer graphene in the four by four representation
reads~\cite{Novoselov_Hall:ref,mccann:086805}
\begin{equation}\label{Ham_bilayer:eq}
    H  = \begin{pmatrix}
            0 & 0 & 0 & v p_- \\
            0 & 0 & v p_+ & 0 \\
            0 & v p_- & 0 & \gamma_1 \\
            v p_+ & 0 & \gamma_1 & 0 \\
          \end{pmatrix},
\end{equation}
where $p_\pm = p_x \pm i p_y$ (note that $p_+ p_- = p_x^2 +p_y^2 = \textbf{p}^2$).
To have a result independent from the choice of the coordinate systems it is convenient to introduce the vector $\textbf{p}= (p_x,p_y,0)$.
Here $v= \sqrt{3} a \gamma_0/(2\hbar)$, where $a$ is the lattice constant in the honeycomb lattice,
$\gamma_0$ is the intralayer coupling between nearest neighbor carbon atoms, while
$\gamma_1$ is the strongest interlayer coupling between two carbon atoms that are on the top of each other.
In the above Hamiltonian we neglect the trigonal
warping~\cite{mccann:086805,koshino:245403,cserti:066802}
since the other interlayer couplings are much smaller than $\gamma_1$.
We have also calculated the time dependence of the position operator when the trigonal warping term is included.
In this case the result is more cumbersome (not presented here) but the structure of the position operator
and the steps of its derivation are similar to that presented below when the trigonal warping is omitted.
Therefore, for the sake of simplicity we now only consider the bilayer graphene without trigonal warping.

In what follows it is convenient to work with dimensionless Hamiltonian.
Thus we re-scale the momentum and the Hamilton operators as
$\tilde{p}_\pm =2 v\, p_\pm/\gamma_1 $, and $\tilde{H}=2 H/\gamma_1$, and
then the Hamiltonian takes the following form (we omit the tilde):
\begin{equation}\label{Ham_bilayer-rescale:eq}
    H  =  \begin{pmatrix}
            0 & 0 & 0 & p_- \\
            0 & 0 & p_+ & 0 \\
            0 & p_- & 0 & 2 \\
            p_+ & 0 & 2 & 0 \\
          \end{pmatrix}.
\end{equation}
To calculate the time dependence of the operator $\textbf{x}(t)$ we shall use Eq.~(\ref{x_op-Q-main-shift:eq}).
After some matrix algebra we found the following result for the decomposition of the Hamilton operator (\ref{Ham_bilayer-rescale:eq}) into a sum of projection operators:  $H  = \sum_{a=1}^4 E_a\, Q_a$, where
\begin{eqnarray}
      \label{H_Q-def:eq}
       E_1 &=& -\Omega -1, \quad E_2 = -\Omega +1, \quad E_3 =\Omega -1, \quad E_4 =\Omega +1, \quad
       \Omega = \sqrt{1 + p_+ p_- } = \sqrt{1+ \textbf{p}^
       2},      \\[2ex]
       Q_1 &=& \frac{I_4 - R}{2}\, \frac{I_4 + T}{2}, \quad
       Q_2 = \frac{I_4 + R}{2}\, \frac{I_4 - T}{2}, \quad
       Q_3 = \frac{I_4 - R}{2}\, \frac{I_4 - T}{2}, \quad
       Q_4 = \frac{I_4 + R}{2}\, \frac{I_4 + T}{2}, \quad         \\[2ex]
       R &=& \left(
               \begin{array}{cccc}
                 0 & e^{-2i\varphi} & 0 & 0 \\
                 e^{2i\varphi} & 0 & 0 & 0 \\
                 0 & 0 & 0 & 1 \\
                 0 & 0 & 1 & 0 \\
               \end{array}
             \right)
        ,  \quad
        T = \frac{1}{\Omega}\left(
              \begin{array}{cccc}
                 -1 & 0 & p_- & 0 \\
                 0 & -1 & 0 & p_+ \\
                 p_+ & 0 & 1 & 0 \\
                 0 & p_- & 0 & 1 \\
              \end{array}
             \right) , \quad \text{and} \quad p_\pm = |\textbf{p}| \, e^{\pm i\, \varphi} ,
\end{eqnarray}
and $I_4$ is the $4 \times 4$ unit matrix.
One can easily show that $R^2 = T^2 = I_4$, ie, both $R$ and $T$ are involutory operators.
Similarly, $R$ and $T$ commute with each other, ie, $\left[ R,T\right]=0$, and the four operators
$Q_1, \dots, Q_4$ are indeed projection operators satisfying the usual relations: $Q_a Q_b = \delta_{a b} \, Q_a $
and $\sum_{a=1}^4 Q_a  = I_4$.
One can also show by direct calculation that $H  = \sum_{a=1}^4 E_a\, Q_a$ holds.

We now can use Eq.~(\ref{x_op-Q-main-shift:eq}) or (\ref{x-sol-exp-H-1:eq})
to find time dependence of the position operator.
Note that to obtain the explicit form of the Zitterbewegung amplitudes it is more useful
to apply Eq.~(\ref{Z_Qa-grad-Qb:eq})
since it is trivial to calculate the derivative of the Hamiltonian (\ref{Ham_bilayer-rescale:eq}) with respect to the momentum operator $\textbf{p}$.
After a tedious but straightforward calculation we have
\begin{widetext}
\begin{eqnarray}
\textbf{x}(t) &=& \textbf{x}(0) +  \textbf{W}\, t  + \textbf{Z}(t),
\quad \text{where} \\
\textbf{W} &=& \sum_{i=1}^4 \frac{\partial E_i}{\partial \textbf{p}}\, Q_i =
\frac{\partial \Omega}{\partial \textbf{p}}\, R T = \frac{\textbf{p}}{\Omega}\, R T
= \frac{\textbf{p}}{\Omega^2}\, \left(
                                \begin{array}{cccc}
                                  0 & -\,\frac{p_-^2}{\textbf{p}^2}  & 0 & p_- \\[1ex]
                                  -\,\frac{p_+^2}{\textbf{p}^2} & 0 & p_+ & 0 \\[1ex]
                                  0 & p_- & 0 & 1 \\[1ex]
                                  p_+ & 0 & 1 & 0 \\[1ex]
                                \end{array}
                              \right)
,
\label{Wt-veg-bilayer:eq}
\end{eqnarray}%
and
\begin{eqnarray}
\textbf{Z}(t)&=&  \frac{\hbar}{4\,\Omega^2}\, \frac{\textbf{e}\times \textbf{p}}{\textbf{p}^2}\,
\left\{
 \left(\cos (2\,\alpha\, t) -1\right)\left(
                                 \begin{array}{cccc}
                                  \beta^2 & 0 & -\beta \,p_- & 0 \\[1ex]
                                  0 & -\beta^2 & 0 & \beta \,p_+ \\[1ex]
                                  -\beta \,p_+ & 0 & \textbf{p}^2 & 0 \\[1ex]
                                  0 & \beta \,p_- & 0 & -\textbf{p}^2 \\[1ex]
                                \end{array}
                              \right)
+  i \sin (2\,\alpha\, t) \left(
                           \begin{array}{cccc}
                             0 & \beta^2 \,\frac{p_-^2}{\textbf{p}^2} & 0 & -\beta \,p_- \\[1ex]
                             -\beta^2 \,\frac{p_+^2}{\textbf{p}^2} & 0 & \beta \,p_+ & 0 \\[1ex]
                             0 & -\beta \,p_- & 0 & \textbf{p}^2 \\[1ex]
                             \beta \,p_+ & 0 & -\textbf{p}^2  & 0 \\[1ex]
                           \end{array}
                         \right)
\right.
\nonumber \\[1.5ex]
 &+& \left.
 \left(\cos (2\,\beta \,t) -1\right)\left(
                                     \begin{array}{cccc}
                                       \alpha^2 & 0 & \alpha \,p_- & 0 \\[1ex]
                                       0 & -\alpha^2 & 0 & -\alpha \,  p_+ \\[1ex]
                                       \alpha \,  p_+ & 0 & \textbf{p}^2 & 0 \\[1ex]
                                       0 & -\alpha \,  p_- & 0 & -\textbf{p}^2 \\[1ex]
                                     \end{array}
                                   \right)
+ i \sin (2\,\beta \, t) \left(
                           \begin{array}{cccc}
                             0 & -\alpha^2 \,\frac{p_-^2}{\textbf{p}^2} & 0 & -\alpha\, p_-   \\[1ex]
                             \alpha^2 \, \frac{p_+^2}{\textbf{p}^2} & 0 & \alpha\, p_+ & 0 \\[1ex]
                             0 & -\alpha\, p_- & 0 & -\textbf{p}^2  \\[1ex]
                             \alpha\, p_+ & 0 & \textbf{p}^2 & 0 \\[1ex]
                           \end{array}
                         \right)
\right.
\nonumber \\[1.5ex]
 &+& \left.
  \left(\cos (2\, t)  -1\right) \left(
                            \begin{array}{cccc}
                                  2\,\textbf{p}^2 & 0 & 2\, p_- & 0 \\[1ex]
                                  0 & -2\,\textbf{p}^2  & 0 & -2\, p_+ \\[1ex]
                                  2\, p_+ & 0 & -2\,\textbf{p}^2 & 0 \\[1ex]
                                  0 & -2\, p_- & 0 & 2\,\textbf{p}^2 \\[1ex]
                                \end{array}
                         \right)
+ i \sin (2\, t) \left(
                                      \begin{array}{cccc}
                                  0 & -2\,p_-^2 & 0 & -2\, p_- \\[1ex]
                                  2\,p_+^2 & 0 & 2\, p_+ & 0 \\[1ex]
                                  0 & -2\, p_- & 0 & 2\,\textbf{p}^2 \\[1ex]
                                  2\, p_+ & 0 & -2\,\textbf{p}^2 & 0 \\[1ex]
                                \end{array}
                                   \right)
\right\}
\nonumber \\[1.5ex]
 &+& \frac{i\, \hbar}{2\, \Omega^3} \, \frac{\textbf{p}}{\textbf{p}^2}
 \left\{
  \left(\cos (2\, \Omega\, t) -1\right) \left(
                           \begin{array}{cccc}
                             0 & 0 & \Omega \,p_- & 0    \\[1ex]
                             0 & 0 & 0 &  \Omega \,p_+ \\[1ex]
                             -\Omega \,p_+ & 0 & 0 & 0 \\[1ex]
                             0 &  -\Omega \,p_- & 0 & 0 \\[1ex]
                           \end{array}
                         \right)
                         +  i \, \sin (2\, \Omega\, t) \left(
                                     \begin{array}{cccc}
                                       0 & -p_-^2 & 0 & -p_-   \\[1ex]
                                       -p_+^2 & 0 & -p_+ & 0 \\[1ex]
                                       0 & -p_- & 0 & \textbf{p}^2  \\[1ex]
                                       -p_+ & 0 & \textbf{p}^2 & 0 \\[1ex]
                                     \end{array}
                                   \right)
 \right\},
\label{Zt-veg-bilayer:eq}
\end{eqnarray}%
\end{widetext}%
where  $\textbf{e}=(0,0,1)$, $\alpha = \Omega-1 , \, \beta=\Omega + 1 $ and $t$ is in units of $\hbar/\gamma_1$.
One can easily see that the above operator $\textbf{x}(t)$ is a hermitian operator.
The vector $\textbf{e}\times \textbf{p}=(-p_y,p_x,0)$ is perpendicular to the momentum vector $\textbf{p}=(p_x,p_y,0)$.
Therefore, the operator $\textbf{x}(t)$ has one longitudinal and three transversal modes parallel and perpendiclar to the momentum $\textbf{p}$, respectively.
It is clear from the result that the oscillatory term $\textbf{Z}(t)$
corresponding to Zitterbewegung is a sum of individual oscillatory terms with four different frequencies.


\begin{thebibliography}{10}

\bibitem{Schrodinger:cikk}
E. Schr\"odinger, Sitzungsber. Preuss. Akad. Wiss. Phys. Math. Kl. {\bf 24},
  418  (1930).

\bibitem{Feshbach_RevModPhys.30.24}
H. Feshbach and F. Villars, Rev. Mod. Phys. {\bf 30},  24  (1958).

\bibitem{Barut_PhysRevD_23_2454}
A.~O. Barut and A.~J. Bracken, Phys. Rev. D {\bf 23},  2454  (1981).

\bibitem{Barut_PhysRevD.31.1386}
A.~O. Barut and W. Thacker, Phys. Rev. D {\bf 31},  1386  (1985).

\bibitem{Cannata_PhysRevB.44.8599}
F. Cannata and L. Ferrari, Phys. Rev. B {\bf 44},  8599  (1991).

\bibitem{PhysRevLett.93.043004}
P. Krekora, Q. Su, and R. Grobe, Phys. Rev. Lett. {\bf 93},  043004  (2004).

\bibitem{Schliemann_Loss_PhysRevLett.94.206801}
J. Schliemann, D. Loss, and R.~M. Westervelt, Phys. Rev. Lett. {\bf 94},
  206801  (2005).

\bibitem{Shen_PhysRevLett.95.187203}
S.-Q. Shen, Phys. Rev. Lett. {\bf 95},  187203  (2005).

\bibitem{Bruder_zitter_PhysRevB.72.045353}
M. Lee and C. Bruder, Phys. Rev. B {\bf 72},  045353  (2005).

\bibitem{Nikolic_PhysRevB.72.075335}
B.~K. Nikoli\'c, L.~P. Z\^arbo, and S. Welack, Phys. Rev. B {\bf 72},  075335
  (2005).

\bibitem{Zawadzki_PhysRevB.72.085217}
W. Zawadzki, Phys. Rev. B {\bf 72},  085217  (2005).

\bibitem{Schliemann:085323}
J. Schliemann, D. Loss, and R.~M. Westervelt, Phys. Rev. B {\bf 73},  085323
  (2006).

\bibitem{brusheim:205307}
P. Brusheim and H.~Q. Xu, Phys. Rev. B {\bf 74},  205307  (2006).

\bibitem{Katsnelson_Zitter_min-cond:cikk}
M.~I. Katsnelson, Eur. Phys. J. B {\bf 51},  157  (2006).

\bibitem{Zawadzki-3:cikk}
T.~M. Rusin and W. Zawadzki, J. Phys.: Condens. Matter {\bf 19},  136219
  (2007).

\bibitem{Schliemann_side-jump_ISI:000243895600066}
J. Schliemann, Phys. Rev. B {\bf 75},  045304  (2007).

\bibitem{PhysRevLett.98.253005}
L. Lamata, J. Le\'on, T. Sch\"atz, and E. Solano, Phys. Rev. Lett. {\bf 98},
  253005  (2007).

\bibitem{PhysRevA.76.041801}
A. Bermudez, M.~A. Martin-Delgado, and E. Solano, Phys. Rev. A {\bf 76},
  041801(R)  (2007).

\bibitem{Winkler_Ulrich_ISI:000246890900071}
R. Winkler, U. Zulicke, and J. Bolte, Phys. Rev. B {\bf 75},  205314  (2007).

\bibitem{Zawadzky_graphene_ISI:000251326800144}
T.~M. Rusin and W. Zawadzki, Phys. Rev. B {\bf 76},  195439  (2007).

\bibitem{Schliemann_Loss_ISI:000248866900038}
E. Bernardes, J. Schliemann, M. Lee, J. C. Egues, D. Loss,
Phys. Rev. Lett. {\bf 99},  076603  (2007).

\bibitem{Zhang_ISI:000254292300028}
X. Zhang, Phys. Rev. Lett. {\bf 100},  113903  (2008).

\bibitem{Vaishnav_ISI:000255117800018}
J.~Y. Vaishnav and C.~W. Clark, Phys. Rev. Lett. {\bf 100},  153002  (2008).

\bibitem{Wang_ISI:000255457100201}
Z.-Y. Wang and C.-D. Xiong, Phys. Rev. A {\bf 77},  045402  (2008).

\bibitem{Schliemann_ISI:000254543000075}
J. Schliemann, Phys. Rev. B {\bf 77},  125303  (2008).

\bibitem{Maksimova_ISI:000259690800083}
V.~Y. Demikhovskii, G.~M. Maksimova, and E.~V. Frolova, Phys. Rev. B {\bf 78},
  115401  (2008).

\bibitem{Loss_SO_ISI:000260574400091}
R.~S. Calsaverini, E. Bernardes, J.~C. Egues, and D. Loss, Phys. Rev. B {\bf
  78},  155313  (2008).

\bibitem{Vertesi_ISI:000261215400050}
R. Englman and T. V\'ertesi, Phys. Rev. B {\bf 78},  205311  (2008).

\bibitem{Maksimova_ISI:000262245400084}
G.~M. Maksimova, V.~Y. Demikhovskii, and E.~V. Frolova, Phys. Rev. B {\bf 78},
  235321  (2008).

\bibitem{Tan_ISI:000262246400065}
S.~G. Tan, M.~B.~A. Jalil, X.-J. Liu, and T. Fujita, Phys. Rev. B {\bf {78}},
  245321  ({2008}).

\bibitem{atomic_Zitter_EPL:cikk}
M. Merkl, F.~E. Zimmer, G. Juzeli$\tilde{\textrm{u}}$nas, and P. \"Ohberg, EPL
  {\bf 83},  54002  (2008).

\bibitem{rusin:045416}
T.~M. Rusin and W. Zawadzki, Phys. Rev. B {\bf 80},  045416  (2009).

\bibitem{Cserti_Zitter-ISI:000242409000014}
J. Cserti and G. David, Phys. Rev. B {\bf 74},  172305  (2006).

\bibitem{Luttinger_PhysRev.102.1030}
J.~M. Luttinger, Phys. Rev. {\bf 102},  1030  (1956).

\bibitem{Murakami_PhysRevB.69.235206}
S. Murakami, N. Nagaosa, and S.-C. Zhang, Phys. Rev. B {\bf 69},  235206
  (2004).

\bibitem{Novoselov_Hall:ref}
K. Novoselov {\it et~al.}, Nature Phys. {\bf {2}},  177  ({2006}).

\bibitem{mccann:086805}
E. McCann and V.~I. Fal'ko, Phys. Rev. Lett. {\bf 96},  086805  (2006).

\bibitem{koshino:245403}
M. Koshino and T. Ando, Phys. Rev. B {\bf 73},  245403  (2006).

\bibitem{cserti:066802}
J. Cserti, A. Csord\'{a}s, and G. D\'{a}vid, Phys. Rev. Lett. {\bf 99},  066802
   (2007).

\bibitem{Berry_phase_120:cikk}
M.~V. Berry, Proc. R. Soc. Lond. A {\bf 392},  45  (1984).

\bibitem{Ross_Nature_463_68_2010.cikk}
R. Gerritsma {\it et~al.}, Nature {\bf 463},  68  (2010).

\end{thebibliography}
\end{document}